\documentclass[twocolumn,preprintnumbers,amsmath,amssymb,prl]{revtex4}
\usepackage{graphicx}% Include figure files
\usepackage{epstopdf}
\usepackage{dcolumn}% Align table columns on decimal point
\usepackage{bm}% bold math
\usepackage{color}

\begin{document}

ACS Nano, accepted (09.03.2017).\\

\title{Comment on ``Spin-Orbit Coupling Induced Gap in Graphene on Pt(111) with Intercalated Pb Monolayer''}

\author{Yuriy Dedkov$^{1}$ and Elena Voloshina$^{2}$}

\affiliation{$^1$Fachbereich Physik, Universit\"at Konstanz, 78457 Konstanz, Germany}
\affiliation{$^2$Humboldt-Universit\"at zu Berlin, Institut f\"ur Chemie, 10099 Berlin, Germany}

\date{\today}

%\begin{abstract}
%The combination of the surface science techniques (STM, XPS, ARPES) and density-functional theory calculations was used to study the decoupling of graphene from Ni(111) by oxygen intercalation. The formation of the antiferromagnetic (AFM) NiO layer at the interface between graphene and ferromagnetic (FM) Ni is found, where graphene protects the underlying AFM/FM sandwich system. It is found that graphene is fully decoupled in this system and strongly $p$-doped via charge transfer with a position of the Dirac point of $(0.69\pm0.02)$\,eV above the Fermi level. Our theoretical analysis confirms all experimental findings, addressing also the interface properties between graphene and AFM NiO.
%\end{abstract}

\maketitle

Recently a paper of Klimovskikh \textit{et al.}~\cite{Klimovskikh:2017jh} was published presenting experimental and theoretical analysis of the graphene/Pb/Pt(111) system. The authors investigate the crystallographic and electronic structure of this graphene-based system by means of LEED, ARPES, and spin-resolved PES of the graphene $\pi$ states in the vicinity of the Dirac point of graphene. The authors of this paper demonstrate that an energy gap of $\approx200$\,meV is opened in the spectral function of graphene directly at the Dirac point of graphene and spin-splitting of $100$\,meV is detected for the upper part of the Dirac cone. On the basis of the spin-resolved photoelectron spectroscopy measurements of the region around the gap the authors claim that these splittings are of a spin-orbit nature and that the observed spin structure confirms the observation of the quantum spin Hall state in graphene, proposed in earlier theoretical works. Here we will show that careful systematic analysis of the experimental data presented in this manuscript is needed and their interpretation require more critical consideration for making such conclusions. Our analysis demonstrates that the proposed effects and interpretations are questionable and require further more careful experiments.

\begin{figure}[b]
\includegraphics[width=\linewidth]{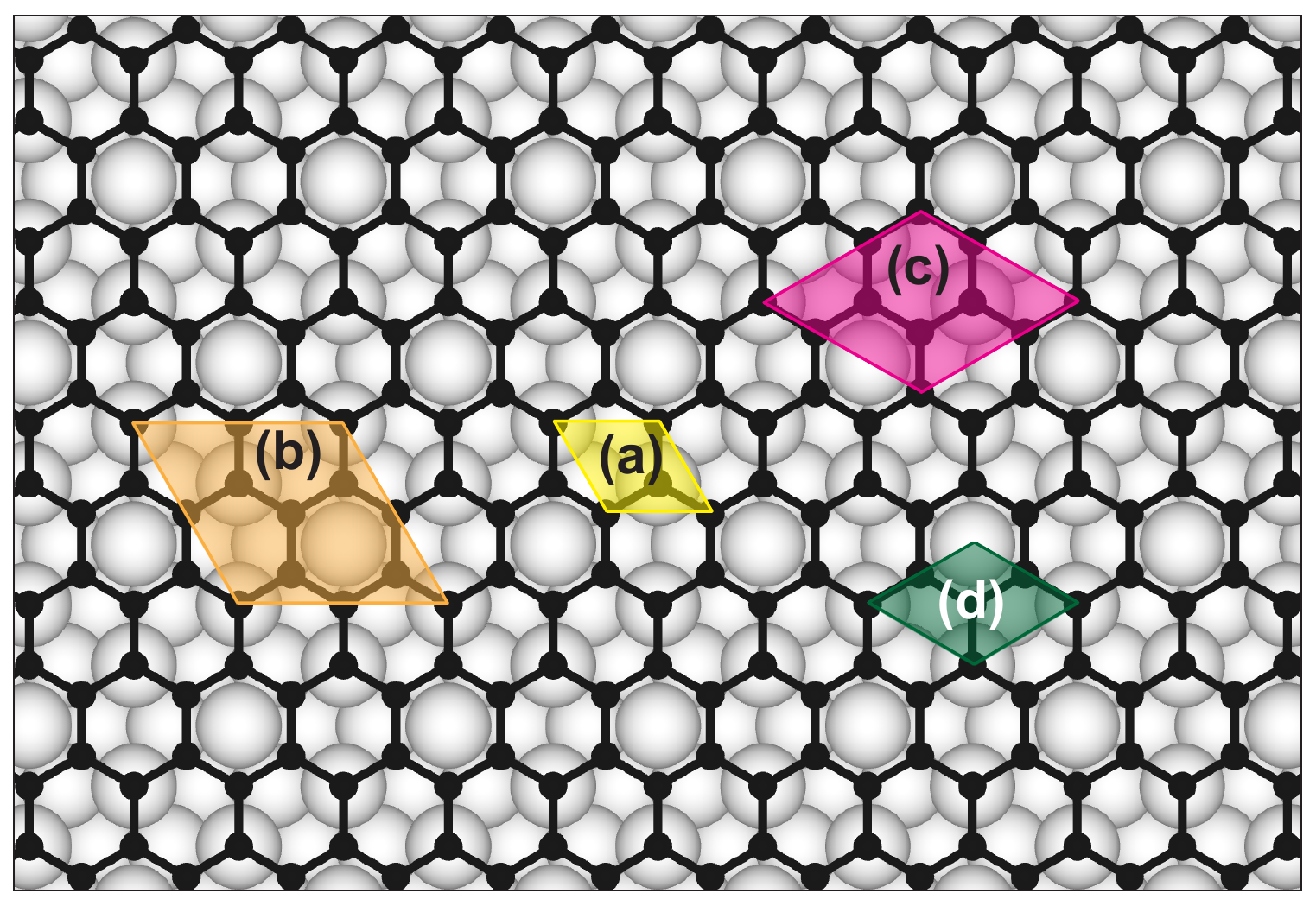}\\
\caption{Identification of different unit cells for the graphene/Pt(111) system: (a) unit cell of $(1\times1)$-graphene, (b) $(2\times2)$-graphene/$(\sqrt{3}\times\sqrt{3})R30^\circ$-Pt(111), (c) $(\sqrt{3}\times\sqrt{3})R30^\circ$-graphene/$(3/2\times3/2)$-Pt(111), and (d) $(2/\sqrt{3}\times2/\sqrt{3})R30^\circ$-graphene/$(1\times1)$-Pt(111).}
\label{structure}                                                                                            
\end{figure}

Our first remark, which, however, does not influence the main result of the paper, relates to the assignment of the crystallographic structure of the graphene/Pt(111) interface (see Figure\,1(c,d) of Ref.~\citenum{Klimovskikh:2017jh} and Fig.~\ref{structure} of the present Comment). Here: (a) unit cell of $(1\times1)$-graphene, (b) $(2\times2)$-graphene/$(\sqrt{3}\times\sqrt{3})R30^\circ$-Pt(111), (c) $(\sqrt{3}\times\sqrt{3})R30^\circ$-graphene/$(3/2\times3/2)$-Pt(111), and (d) $(2/\sqrt{3}\times2/\sqrt{3})R30^\circ$-graphene/$(1\times1)$-Pt(111). It is obvious that the assignment made in Ref.~\citenum{Klimovskikh:2017jh} for the structure as $(\sqrt{3}\times\sqrt{3})R30^\circ$-graphene/Pt(111) is incorrect. As was shown in Ref.~\citenum{Voloshina:2011NJP} the correct notation is either (b) or based on the description (d), when $(1\times1)$-Pt(111) layer forms a structure underneath graphene with symmetry $(2/\sqrt{3}\times2/\sqrt{3})R30^\circ$ with respect to the $(1\times1)$-graphene unit cell.

The second critical comment, which is more serious, regards to the spin-resolved photoemission data presented in Ref.~\citenum{Klimovskikh:2017jh}. Our criticism is devoted to the treatment of the presented data and our analysis demonstrates that, generally, the spin-resolved data have to be analyzed with very high accuracy. Furthermore, the declared in Ref.~\citenum{Klimovskikh:2017jh} effect, namely the gap opening at the graphene Dirac point, cannot be related to the spin-orbit induced effects if spin-resolved data presented in the original manuscript are considered.

\begin{figure*}[t]
\includegraphics[width=0.85\textwidth]{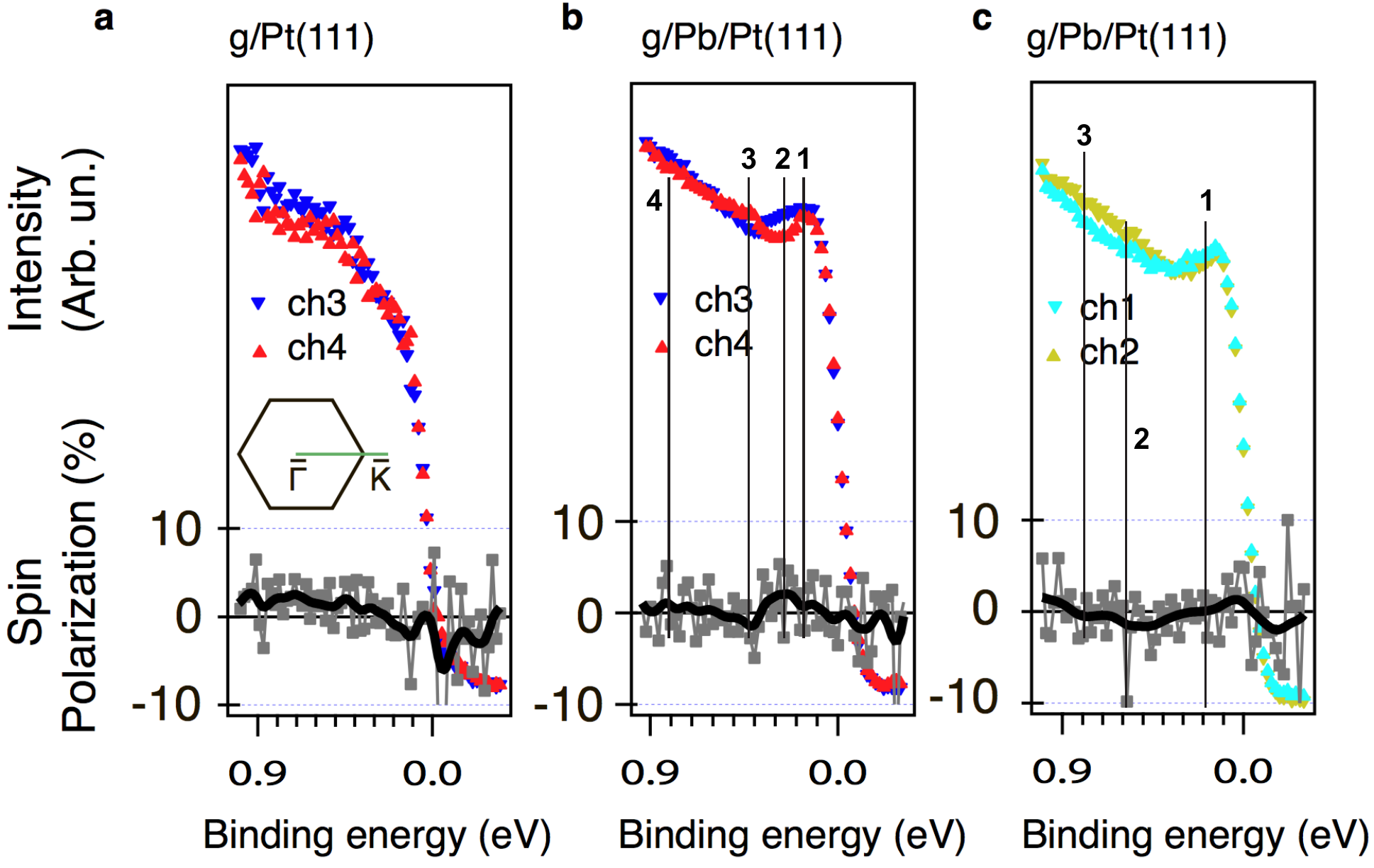}\\
\caption{Data extracted from Fig.\,3 and Fig.\,S5 of the discussed Ref.~\citenum{Klimovskikh:2017jh}. Upper part of every panel shows intensities for the spin-up and spin-down channel and lower part shows the respective spin-polarization. Numbers ($1-4$) and ($1-3$) in panels (b) and (c), respectively, shows the questionable regions of the spectra, which are discussed in the present comment.}
\label{SPPES_spectra}                                                                                            
\end{figure*}

Spin-resolved photoemission experiments allow to separate the spin-integrated signal, which comes after energy analyzer on to channels -- spin-up ($I_{up}$) and spin-down ($I_{down}$). Such experiments are usually performed with the so-called spin-detectors and one of such devices, namely the mini-Mott spin-detector, was used in the discussed work. In the mini-Mott spin-detector the electron beam, which enters the detector, is accelerated to high energies ($26$\,keV) and then it is scattered on the Th-target. In this case, due to the spin-orbit effect, the scattering potential of the Th-atom cannot be considered as symmetric in the space and left/right spin asymmetry with respect to the scattering plane appears (scattering plane is formed by the direction of the incoming on the target the electron beam and the spin quantization axis). Then the back-scattered (scattering angle is $120^\circ$) electrons are detected from the left ($I_L$) and right ($I_R$) channeltrons. Due to the fact that the scattering efficiency for the back-scattered electrons is very low and never reaches $100$\%, the efficiency of the spin-detector is characterized by the so-called Sherman function ($S$), which was set to $0.12$ in Ref.~\citenum{Klimovskikh:2017jh}. The value of the Sherman function for the particular experimental assembly is obtained in the independent experiment on the well-known sample or with the electron beam of known spin polarization. The measured intensities are used for the calculation of the spin-asymmetry $A=(I_L-I_R)/(I_L+I_R)$, spin polarization $P=(I_{up}-I_{down})/(I_{up}+I_{down})=A/S=1/S\cdot(I_L-I_R)/(I_L+I_R)$, and intensities of the spin-up and spin-down channels $I_{up,down}=I_0\cdot(1\pm P)/2$, where $I_0=I_L+I_R$. Here we would like to emphasize that $P$ and $I_{up,down}$ are calculated on the same step. The statistical errors for the spin polarization and spin intensities are $\Delta P=1/(S\sqrt{I_0})$ and $\Delta I_{up,down}=I_{up,down}\cdot\Delta P/(1\pm P)$, respectively~\cite{Johnson:1992}. In the modern spin-resolved photoemission experiments, the spin polarization usually has an error of $2-3$\%.

The main measurement error comes from the experiment-equipment induced spin asymmetry, which is determined by the geometry and construction of the spin-detector assembly, efficiency of the channeltrons, etc.~\cite{Johnson:1992} This effect of the experiment-equipment induced spin asymmetry can be eliminated for magnetic materials via two independent measurements, performed at opposite magnetization directions ($M^+, M^-$)~\cite{Johnson:1992}. For the non-magnetic samples, where spin-orbit effects are investigated, such measurements have to be performed for two opposite helicities of light ($\sigma^+, \sigma^-$)~\cite{Eyers:1984kq,Sconhense:1986ik,Sinkovic:1997hz}. Taking into account this consideration, the final equation for the spin polarization is $P=(1/S)(\sqrt{I^+_L I^-_R}-\sqrt{I^+_R I^-_L})/(\sqrt{I^+_L I^-_R}+\sqrt{I^+_R I^-_L})$ with the corresponding formulas for the spin-up and spin-down channels~\cite{Johnson:1992}. Here the upper signs ``$+$'' and ``$-$'' correspond to the opposite sample magnetizations or light helicities directions. It is obvious that such measurements allow to determine the absolute value of the experiment-equipment induced spin asymmetry and this approach was used for the determination of the setup-induced spin asymmetry via experiments on the magnetic sample as stated by the authors in section Methods of Ref.~\citenum{Klimovskikh:2017jh}. Therefore we can conclude that the spin polarizations presented in Fig.\,3 and Fig.\,S5 of Ref.~\citenum{Klimovskikh:2017jh} as grey points connected by the lines are absolute values with the error of $2-3$\%. The respective figures are compiled in Fig.~\ref{SPPES_spectra} of the present comment.

Before further discussion we would like to emphasise, that according to the previous consideration, the spin polarization value and intensities of the spin-up and spin-down channels are calculated as one step. That means that the respective sign of the spin polarization has to be reflected in the spin-resolved spectra. Analysis of the region of spin-polarization spectra around point (2) in Fig.~\ref{SPPES_spectra} (b) clearly demonstrates that point marked by the vertical line has negative spin polarization, whereas two neighbouring points to the left and to the right have positive polarizations. At the same time considering the spin-resolved spectra presented in the upper panel, we can see that all points have to have the same spin polarization. Therefore there is a clear strong contradiction between two sets of points. Several questionable regions of the spin-resolved spectra together with the respective spin polarization are marked by different numbers in Fig.~\ref{SPPES_spectra} (b) and (c), where the sign of spin polarization $P$ and the respective sign of $(I_{up}-I_{down})/(I_{up}+I_{down})$ extracted from the spectra are opposite. [Here we also would like to point to the confusing caption of Fig.\,3 in Ref.~\citenum{Klimovskikh:2017jh}, namely to the sentence: ``Also presented in (a) is the spin polarization calculated as the raw asymmetry $(I_{up} - I_{down})/(I_{up} - I_{down})$ multiplied by the Sherman function $S_{eff} = 0.12.$'' In this case it is not clear -- what is exactly drawn as a lower part of every panel in (a): spin polarization ($P$) as marked in the axis title, raw asymmetry ($A=(I_L-I_R)/(I_L+I_R)$) as stated in the first part of the quoted sentence, or spin polarization multiplied by $0.12$, i.\,e. the spin asymmetry ($A=SP=S\cdot(I_{up} - I_{down})/(I_{up} - I_{down})=(I_L-I_R)/(I_L+I_R)$) as stated in the second part of the quoted sentence?] In our further analysis we assume that the spin polarization ($P$) is drawn in Fig.\,3 of Ref.~\citenum{Klimovskikh:2017jh}. 

From the close analysis of the lower parts of all panels in Fig.~\ref{SPPES_spectra} we can conclude that the black curve ($P_{av}$), which supposedly is the average curve for the experimental points of $P$ over, at least, 6 experimental points, was used for the calculation of intensities of spin-up and spin-down channels (here we would like to emphasise that, obviously, this curve is not a result of the fitting routine described in Ref.~\citenum{Marchenko:2013iu}). However this is an absolutely inappropriate step for the treatment of the spin-resolved photoemission data. Such average procedure for the spin polarization leads to the dramatic increase of the measurement error from $\Delta P$ to $n\Delta P$ (pessimistic estimation) or to $\sqrt{n}\Delta P$ (optimistic estimation), where $n$ is the number of points used during averaging procedure (in the presented case $n\geq6$). Taking into account the strong scattering of experimental points presented in Fig.~\ref{SPPES_spectra} (grey points in every lower panel), we can conclude that the experimental error for spin polarization is, at least, $\Delta P\approx3$\%. Therefore the resulting error for the averaged black curve will be between $\Delta P_{av}\approx7$\% and $\Delta P_{av}\approx18$\%. The maximal value of the averaged spin polarization is $3$\% (see black curve in Fig.~\ref{SPPES_spectra}(b)) and therefore we have to write the final value as between $P_{av}=(3\pm7)$\% and $P_{av}=(3\pm18)$\% (depending on the estimation method for the error) and the intensities for the spin-up and spin-down channels have to be presented with the respective error bars as well. This analysis clearly demonstrates that the effect offered in the abstract of Ref.~\citenum{Klimovskikh:2017jh} is not supported by the experimental results after proper analysis of the data and statistical analysis. Here we can also conclude that the spin-splitting of $100$\,meV presented in Fig.\,3(a) of Ref.~\citenum{Klimovskikh:2017jh} does not exist or it has to be much smaller. This value can be only estimated after proper analysis of the experimental data or after new spin-resolved experiments.

In conclusion, we demonstrated that the analysis and presentation of the spin-resolved experimental results given in Ref.~\citenum{Klimovskikh:2017jh} is very confusing and the presented data require more careful analysis and interpretation. Our discussion shows that the appropriate error analysis is required during such spin-resolved photoemission experiments, because in some cases it might lead to the dramatic influence on the raw data. Our careful investigation of the experimental data presented in the discussed work shows that the main conclusions of the paper on the spin-orbit induced gap opening in graphene are not supported by the data and further accurate spin-resolved investigations of the studied system are necessary.

%\bibliography{/Users/YuDedkov/Work/Articles/___REFERENCES___/references_all.bib}

\begin{thebibliography}{10}

\bibitem{Klimovskikh:2017jh}
Klimovskikh,~I.~I.; Otrokov,~M.~M.; Voroshnin,~V.~Y.; Sostina,~D.;
  Petaccia,~L.; Di~Santo,~G.; Thakur,~S.; Chulkov,~E.~V.; Shikin,~A.~M. Spin-Orbit Coupling Induced Gap in Graphene on Pt(111) with Intercalated Pb Monolayer.
  \textit{ACS Nano} \textbf{2017}, \textit{11}, 368--374.
  
\bibitem{Voloshina:2011NJP}
Voloshina,~E.~N.; Generalov,~A.; Weser,~M.; B{\"o}ttcher,~S.; Horn,~K.;
  Dedkov,~Y.~S. Structural and Electronic Properties of the Graphene/Al/Ni(111) Intercalation System. \textit{New J. Phys.} \textbf{2011}, \textit{13}, 113028.

\bibitem{Johnson:1992}
Johnson,~P.~D.; Brookes,~N.~B.; Hulbert,~S.~L.; Klaffky,~R.; Clarke,~A.;
  Sinkovi~c,~B.; Smith,~N.~V.; Celotta,~R.; Kelly,~M.~H.; Pierce,~D.~T.;
  Scheinfein,~M.~R.; Waclawski,~B.~J.; Howells,~M.~R. Spin-Polarized Photoemission Spectroscopy of Magnetic Surfaces Using Undulator Radiation. \textit{Rev. Sci. Instrum.}
  \textbf{1992}, \textit{63}, 1902--1908.
  
\bibitem{Eyers:1984kq}
Eyers,~A.; Sch\"afers,~F.; Sch\"onhense,~G.; Heinzmann,~U.; Oepen,~H.~P.;
  H\"unlich,~K.; Kirschner,~J.; Borstel,~G. Characterization of Symmetry Properties of Pt(111) Electron Bands by Means of Angle-, Energy-, and Spin-Resolved Photoemission with Circularly Polarized Synchrotron Radiation. \textit{Phys. Rev. Lett.} \textbf{1984}, \textit{52},
  1559--1562.

\bibitem{Sconhense:1986ik}
Sch{\"o}nhense,~G. Photoelectron Spin-Polarization Spectroscopy: A New Method in Adsorbate Physics. \textit{Appl. Phys. A} \textbf{1986},
  \textit{41}, 39--60.

\bibitem{Sinkovic:1997hz}
Sinkovic,~B.; Tjeng,~L.~H.; Brookes,~N.~B.; Goedkoop,~J.~B.; Hesper,~R.;
  Pellegrin,~E.; de~Groot,~F. M.~F.; Altieri,~S.; Hulbert,~S.~L.; Shekel,~E.;
  Sawatzky,~G.~A. Local Electronic and Magnetic Structure of Ni below and above $T_C$: A Spin-Resolved Circularly Polarized Resonant Photoemission Study. \textit{Phys. Rev. Lett.} \textbf{1997}, \textit{79},
  3510--3513.

\bibitem{Marchenko:2013iu}
Marchenko,~D.; Sanchez-Barriga,~J.; Scholz,~M.~R.; Rader,~O.; Varykhalov,~A. Spin Splitting of Dirac Fermions in Aligned and Rotated Graphene on Ir(111).
  \textit{Phys. Rev. B} \textbf{2013}, \textit{87}, 115426.

\end{thebibliography}

This document is the unedited Author's version of a Submitted Work that was subsequently accepted for publication in ACS Nano, copyright \textcopyright\, American Chemical Society after peer review. To access the final edited and published work see doi: 10.1021/acsnano.7b00737.

\end{document}